# Using Generative Artificial Intelligence Creatively in the Classroom and Research: Examples and Lessons Learned

Maria J. Molina,[a] Amy McGovern,[b] Jhayron S. Perez-Carrasquilla,[a] Xiaowen Li,[c] and Robin L. Tanamachi[d]

[a]University of Maryland, College Park, MD

[b]The University of Oklahoma, Norman, OK

[c]Morgan State University, Baltimore, MD

[d]Purdue University, West Lafayette, IN

Corresponding Author: Maria J. Molina, mjmolina@umd.edu

ABSTRACT: Although generative artificial intelligence (AI) is not new, recent technological breakthroughs have transformed its capabilities across many domains. These changes necessitate new attention from educators and specialized training within the atmospheric and related sciences. Enabling students to use generative AI effectively, responsibly, and ethically is crucial for their academic and professional development. Educators can also use generative AI to develop engaging classroom activities, such as active learning modules and games; however, they must be aware of potential pitfalls and biases. There are also ethical implications in using tools that lack transparency and have a considerable carbon footprint, as well as equity concerns for students who lack access to more sophisticated paid versions of generative AI tools and have deficiencies in prior educational training. This article is written for students and educators alike, particularly those interested in learning more about generative AI in education and research, including its use cases, ethical concerns, and a brief history of its emergence. Sample user prompts are also provided across numerous applications in education and the atmospheric and related sciences. Current solutions addressing broader ethical concerns regarding the use of generative AI in education remain limited; however, this work aims to foster a discussion that could galvanize the education community around shared goals and values.

SIGNIFICANCE STATEMENT: Recent technological advances have transformed intelligent machines into much more useful tools across a wide range of applications, leading to rapid career changes. However, few educators and students are familiar with the history of these technologies, nor do they have concrete examples of how to apply them in the atmospheric and related sciences. We aim to fill this gap by providing a brief history of its emergence and examples of the uses of intelligent machines, including lesson planning, literacy training, concept understanding, and software development. Examples of subtle errors produced by intelligent machines are also included, illustrating that domain expertise is necessary when using these technologies, given the far-reaching societal impacts of the atmospheric and related sciences.

CAPSULE: Use examples, limitations, and ethical concerns surrounding generative artificial intelligence in academic environments for the atmospheric and related sciences are discussed.

## 1. AI-driven evolution in education

Technological breakthroughs in generative artificial intelligence (AI) have transformed the tools available to students and faculty in the classroom as well as in most academic environments (McMurtrie 2024). Generative AI is an assistive technology that can generate content across different modalities, including text, software, audio, images, and videos. All these uses make it a powerful tool for education and research. However, new opportunities provided by generative AI also introduce new complexity. For example, how can educators use generative AI responsibly and enable students to use these tools in the classroom and their careers?

Although the use of generative AI in education is still in its early stages, initial studies suggest that it offers pedagogical benefits. For example, some studies show that pairing generative AI with other tools in math, programming, and engineering educational settings increases task-completion rates, enhances students' learning experiences by making the interaction with technical tools easier, increases student interest in the subject matter, and facilitates cross-disciplinary learning (Gouia-Zarrad and Gunn 2024; Liu and Yang 2024; Nam et al. 2024; Matzakos and Moundridou 2025). In math courses, early studies indicate that tutoring from generative AI yields learning gains equivalent to human-based tutoring (Pardos and Bhandari 2024). The time saved in creating academic content with generative AI could also be used to expand critical thinking and delve deeper into complex questions in the classroom (Bull and Kharrufa 2023). Other studies highlight adverse pedagogical outcomes, including hindering the development of creativity and problem-solving skills (Zhai et al. 2024). Using generative AI when writing can also result in weaker and less distributed brain activity, along with reduced memory recall and less self-reported ownership of authorship (Kosmyna et al. 2025). Further guidance is needed on how to effectively use generative AI in education and research, particularly in atmospheric and related sciences.

## 2. A very brief history of generative AI

One of the first chatbots (formerly "chatterbots") was developed at a university in the 1960s (Figure 1) and named ELIZA (Weizenbaum 1966). ELIZA could be considered a generative AI model, but more aptly, it is a natural language processing program that returns predefined responses using pattern matching and text substitution. Components of "backpropagation" were introduced in the 1960s, referring to the backward propagation of prediction error through a neural network. Formal backpropagation was introduced in the 1980s (Rumelhart et al. 1986) and involves computing the gradient of the loss function with respect to the network's parameters (weights and biases) to update the parameters and reduce error (i.e., "learning"). Backpropagation is still a key concept in modern AI.

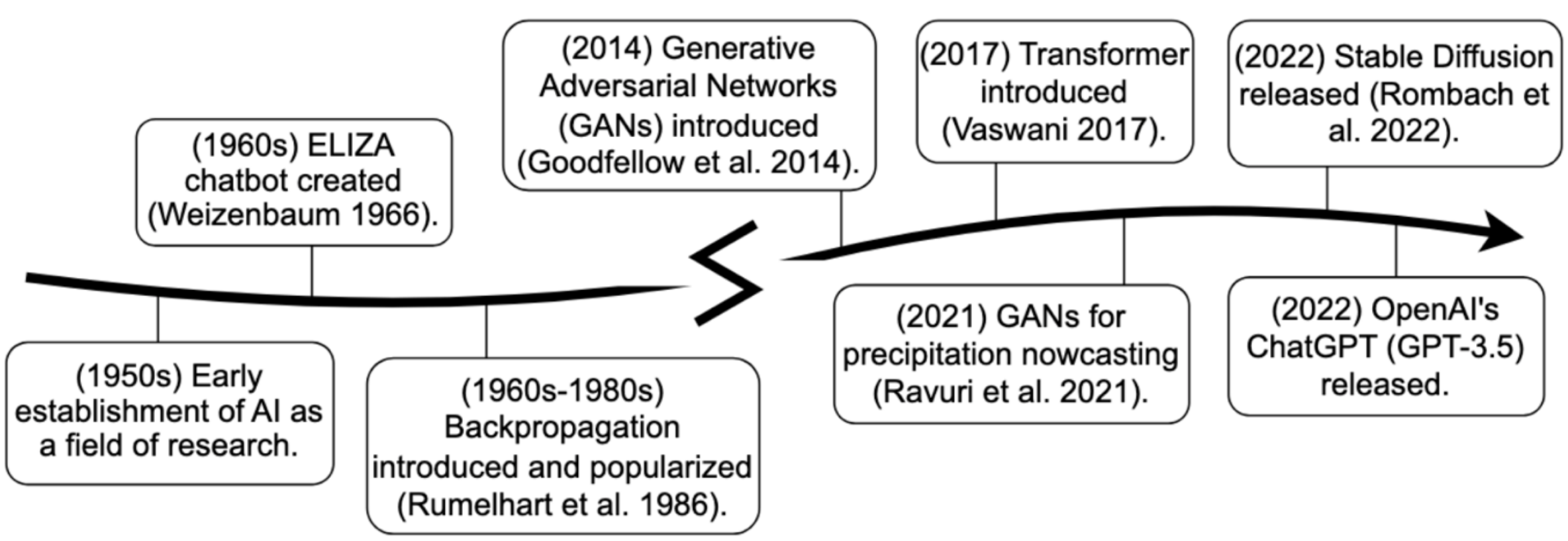

Figure 1. Brief history of generative artificial intelligence (AI). Note that other notable advances in machine learning more broadly occurred between the 1970s and 2000s.

Virtual assistants gained popularity during the 2010s due to their successful responses, which utilized natural language processing in both audio and text forms on our smartphones and other devices. Generative AI gained considerable traction within the computer science field in 2014 with the development of Generative Adversarial Networks (GANs; Goodfellow et al. 2014); a learning framework where one neural network generates increasingly realistic content in an attempt to fool another neural network that discriminates the realism of the generated content. Generative AI is distinct from traditional AI in that it models the underlying data structure and can thus generate new data by sampling from its learned distribution. Since then, GANs have seen successful implementations in atmospheric and related sciences for tasks such as predicting and downscaling (see Molina et al. 2023, for a review). A notable early use of GANs for precipitation nowcasting was reviewed favorably by expert meteorologists due to skillful forecasts with limited "blurring," a notable limitation with other deep learning methods at the time (Ravuri et al. 2021). Following these advances, the release of Stable Diffusion (Rombach et al. 2022) in August 2022 (used in AlphaFold-3; Abramson et al. 2024) and OpenAI's ChatGPT (Generative Pre-trained Transformer) in November 2022 (version 3.5) sparked widespread public interest in generative AI, followed by the launch of other popular platforms, such as Google Gemini, Microsoft Copilot, and Meta Llama.

The backbone of ChatGPT is a Transformer-style architecture (Vaswani et al. 2017) developed to overcome the limitations of previous AI models that used sequential data. AI models that processed sequential data relied on recurrent and convolutional neural networks to encode sequences (e.g., natural language) and decode output(s), limiting learning to sequential computation. Attention mechanisms enable parallel computing by considering the whole input sequence, allowing a model to focus on the most relevant parts of the sequence[1] that provide more context and relevance to predictions (Vaswani et al. 2017). While attention connects different sequences, "self-attention" connects elements within the same sequence and is used by the Transformer. Transformers are a flexible architecture that can be generative, depending on the defined learning and inference tasks.

As described in OpenAI's GPT-4 technical report (Achiam et al. 2023), publicly available data and licensed data from third-party developers were used to pre-train the Transformer-style model to predict the next "token" (i.e., encoded word) in a document. The model was then fine-tuned with human input (i.e., Reinforcement Learning from Human Feedback). Interactions throughout a session (i.e., "conversation") with ChatGPT provide context for subsequent responses, enabling interactivity and refinement of output based on human-provided prompts. In addition to text generators (e.g., based on Transformer-style models) and image generators (e.g., based on diffusion-style models; Rombach et al. 2022), software generation assistants are also becoming more commonplace (Puryear and Sprint 2022). Other generative AI models span multiple modalities, supporting a range of uses for educators and students.

---

[1] Mathematically, by adding weights.

## 3. Generative AI uses for educators

### *a. Generative AI strengths*

Educators can use generative AI in many ways (Table 1; prompt responses in supplementary material). Generative AI tools can accelerate lesson planning, particularly for junior educators or educators creating a new course. For lesson planning, providing the generative AI with sufficient background information can help guide its output to meet the course's needs and increase the efficiency of its use. These details can include the educational level of students (e.g., undergraduate or graduate), course learning outcomes, class size, class duration, and instructional or learning environments (e.g., laboratory sessions, field experiments, or online courses). The interactivity of some generative AI models allows for the refinement of generated content. For example, suppose the AI suggests an active learning activity that requires too much movement around classroom furniture. In that case, an educator can prompt the AI for a less movement-intensive activity to better suit the classroom layout and ensure student safety.

Educators can also provide their existing lesson plans to the generative AI tool and then request updated versions that integrate active learning techniques (e.g., think-pair-share or real-world case studies) and consider various student learning modalities (Nancekivell et al. 2020). More recently, generative AI ‘agents’ can be created, which fine-tune existing generative AI based on course content (e.g., Copilot Agents). Generative AI also excels in the gamification of lessons and the creation of scripts for role-playing scenarios; both may be subsequently heavily edited by the educator, but the AI-generated content can serve as an inspirational starter. For university principal investigators, generative AI can assist with creating project or group acronyms (Table 1). Generative AI can also incorporate specific types of humor when prompted, making research, learning, and classroom activities more fun and engaging.

Table 1. Sample generative AI user prompts. Categories are “lesson planning” (design of curricula, lecture content, and educational materials), “assessment design” (creation of quiz and exam questions and formative assessment tools), “student engagement” (applications that enhance

classroom interactivity, gamification, and active learning strategies), and "AI literacy training" (scenarios where AI is used to teach students about AI itself, including critical evaluation of AI-generated content and ethical considerations).

| Educator Use Category | Example User Prompts |
|---|---|
| Lesson Planning | "Help me create a lesson plan for the first day of class that is 50 minutes in length, that includes the instructor sharing the class syllabus, undergraduate students introducing themselves, and an active learning activity to assess students' baseline knowledge of the course content." |
| Lesson Planning | "Create a lesson plan for an undergraduate class of 24 students 75 minutes in length that addresses the following learning outcome: 'Explain the processes affecting radiative transfer in the atmosphere and global energy balance'. Provide a formative and summative assessment." |
| Lesson Planning | "For a senior undergraduate introductory class in AI, I need an idea for an in-class exercise to help them better understand neural networks." |
| Assessment Design | "I'm teaching an overview course in artificial intelligence using the Russell and Norvig book. I am looking for ideas about quiz questions on choosing appropriate state spaces for reinforcement learning." |
| Assessment Design | "Give me some quiz ideas for Monte Carlo tree search." |
| Assessment Design | "For a junior-level mesoscale meteorology class, I need an example case of a mesoscale convective system (MCS) for students to analyze. Please formulate an assignment whose main objective is to diagnose the primary drivers of the MCS using the convective storm matrix (which I will cover in class), and whose primary artifact is an extended abstract style write-up including hand-drawn and computational analyses." |
| Assessment Design | "Help me create a homework assignment for senior-level undergraduates that assesses their understanding of atmospheric vertical structure and composition." *follow up* "This assignment is too long. Can you make it so the written report would only be 2-3 pages long?" |
| Student Engagement | "Can you explain to me like I am 10 years old how geostationary satellite data is put on a map?" *follow up* "What is 'georeferencing'?" |
| Student Engagement | "Help me come up with mnemonics for undergraduate students that help them understand skew-T terms for vertical profiling of the atmosphere." |
| Student Engagement | "Can you help me make an acronym for PARETO that involves climate, weather extremes, and machine learning?" *follow up* "Can you make one that is funny to graduate students?" |
| Student Engagement | "Can you help me come up with an acronym for a field study that involves observing urban heat at the street level?" |
| Student Engagement | "Can you please help me summarize the following notes on our research discussion?" *follow up* "Can you please instead summarize these notes organized by 'grand challenges,' 'gaps,' and 'opportunities'?" |
| Student Engagement | "Help me summarize my lecture notes into three key take-away statements for students." |
| AI Literacy Training | (Instructor prompt to students): "Ask an AI chatbot the following question about the role of water in the Earth's atmosphere: ... Evaluate the accuracy of its responses relative to what we learned in class." |
| AI Literacy Training | (Instructor prompt to students): "Ask an AI to generate an illustration of a meteorologist at work. What are some assumptions made of their employment?" |

For course assignments and exams, generative AI can accelerate the creation of questions. The educator should provide sufficient guidance to the AI, including details on question format (e.g., multiple-choice or single-choice) and accompanying answer key. Generative AI also excels as a tutor, which can help prepare educators to answer questions from students with a broad array of learning modalities and backgrounds, given that a deep understanding of content is crucial for successfully answering student questions. Educators may also "flip the script" and engage (or have students engage) with the generative AI tool as if it were a virtual student.

Educators should consider whether incorporating generative AI tools into course assignments aligns with course outcomes and student preparation, taking into account the implications of generative AI in the real world and its impact on students' careers. Critically assessing content produced by generative AI and being literate about its benefits and pitfalls are increasingly essential skills in today's workforce. Ways to incorporate the use of generative AI in assignments include having a student use generative AI to create an essay and then having the same student mark up the produced essay, or having students use generative AI to create illustrations for a class presentation. Students will inevitably use AI tools in their education and careers; therefore, incorporating generative AI into assignments may be a worthwhile investment for educators, provided students engage critically and uphold academic integrity when using generative AI.

*b. Generative AI limitations*

While they are powerful computational assistants, generative AI tools have numerous limitations. For instance, generative AI may sometimes hallucinate and include details that appear to be factual but are fabricated. Figures containing scientific data should not be uploaded for direct modification because the generative AI version may contain fabricated data points. Content of other modalities, like text, can also include subtle errors that an educator may easily overlook (Figure 2). Thus, generative AI should not be used as a search engine, and we caution educators against using generated content directly. The use of generative AI also makes it difficult, in most cases, to identify the sources of generated ideas, some of which may have been reproduced to the extent that they could be considered plagiarized. Students may also be able to easily find answers to AI-generated questions and assignments, as the training data includes publicly available datasets. In our experiences, we would use the suggested questions to help create new questions we would not have thought of on our own. Moreover, the perceived quality of the AI-generated content will depend on the quality of the user prompts, which may take practice to refine. Educators should offer students opportunities to *critique* the output of generative AI tools, particularly relative to content that they have recently mastered.

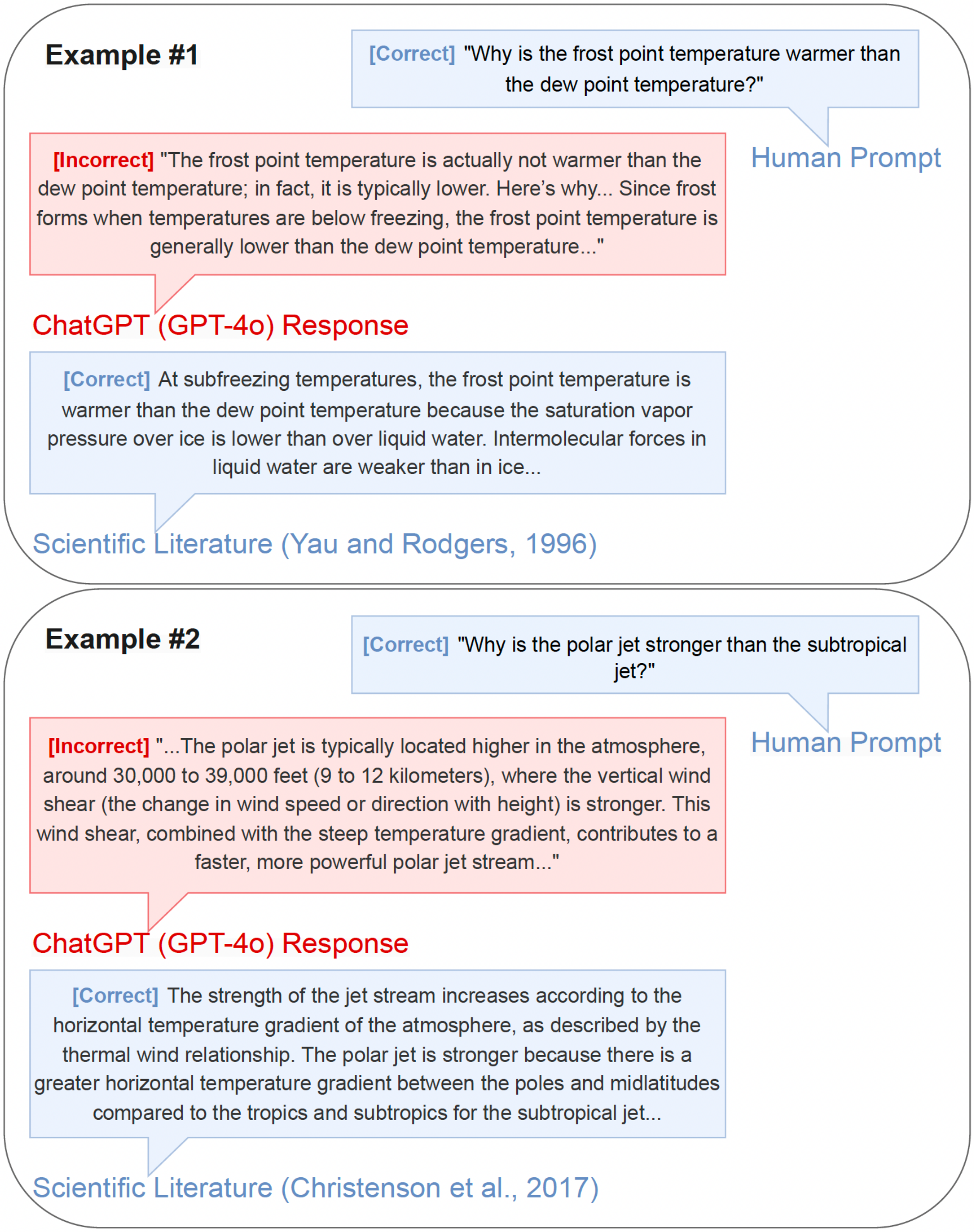

Figure 2. Examples of **incorrect** generative AI responses (red; generated by Open AI's GPT-4o in August 2024) to **correct** human prompts (blue) and a **correct** response based on scientific literature (blue). For example #1, the frost point temperature is indeed higher than the dew point at subfreezing temperatures, an important thermodynamic property for frost formation (Yau and Rogers 1996). For example #2, the polar jet stream is at a lower altitude (about 9-12 km) than the subtropical jet stream (about 10-16 km), related to the tropopause altitude in the tropics and subtropics (Christenson et al. 2017).

## 4. Generative AI uses for students

### *a. Generative AI strengths*

Generative AI has the potential to assist students in completing routine tasks that involve acquiring knowledge, writing, and developing software (Table \ref{t2}; prompt responses in supplementary material). For instance, in the case of personalized tutoring, generative AI can be prompted to explain concepts and follow a specific tone or teaching style, all of which can be aligned with the student's learning preferences. An advantage of using generative AI for learning is that it is an infinitely patient tool, allowing students to ask for help repeatedly without fear of judgment. Generative AI can also help create templates for emails or letters, which can be very time-consuming for students during the early stages of their professional training. The iterative capabilities of some generative AI tools also enable the creation of resumes or curricula vitae, documents that can help students pursue careers. Generative AI can also help accelerate software development, particularly in tasks that are less desirable or repetitive, such as processing data in specified formats, computing basic statistics, or making aesthetic design changes to figures. Apart from coding specific functions from scratch, generative AI can assist with commenting (e.g., docstrings) and optimizing previously created software, thereby enhancing teamwork and open science practices by making software more understandable and transparent, and improving work reproducibility.

Table 2. Sample generative AI user prompts. Categories are “concept understanding” (interactive tutoring, literature search, and role playing), “writing assistance” (task completion, text summarization, and grammar, tone, and language suggestions), “illustrator” (concept illustration or figure modification), and “software development” (data processing and analysis, language translation, human-readable documentation, and optimization).

| Student Use Category | Example User Prompts |
|---|---|
| Concept Understanding | "Why is the height of the troposphere higher near the equator and tropics, and then lower near the poles? Explain using concepts from the ideal gas law, the hypsometric equation, and the latitudinal variation to incident radiation." |
| Concept Understanding | "Why does the hydrostatic equation matter?" |
| Concept Understanding | "Please pretend to be a first-year student studying atmospheric science. Ask me a series of questions about atmospheric forces and their role in creating weather. Then, abandon the pretense and evaluate how well I did at answering those questions." |
| Writing Assistance | "I would like to create a one-page resume highlighting my coursework and relevant research experience for a private sector job in hydrology. Please ask me a series of questions about these items and then compose the resume." |
| Writing Assistance | "Please help me draft an introductory cover letter to a prospective graduate adviser." |
| Writing Assistance | "Please help me draft a thank you email to a prospective employer. My interview was earlier today, and it went well. The job opening is for an associate scientist role." |
| Writing Assistance | "What are different ways to say that 'something needs to be addressed in a specific domain'?" |
| Writing Assistance | "Can you please help me rewrite the following sentence to sound less 'harsh'?" (e.g., response or rebuttal to a peer review comment) |
| Illustrator | "Can you help me create a decision tree diagram that shows how someone would pick a machine-learning model for their research? Start from supervised, unsupervised, and reinforcement learning. End with specific model types." |
| Software Development | "Please give me a function in Python to compute the trend of a yearly time series and assess its statistical significance. The time series has a Pandas DataFrame format, and I want to be able to use different types of statistical tests for the significance computation. Please describe what each line of code does." |
| Software Development | "I am performing the computation below on a DataArray with shape (x, y, z), but the code is taking too long to run. Can you help me optimize the code so it runs faster? I could use parallelization since I have several CPUs available." |
| Software Development | "Please add comments to this block of code describing exactly what the code is doing." |
| Software Development | "Please create a docstring for the following function that describes the data type of the input arguments and describes what the function does." |
| Software Development | "Convert this Python code into FORTRAN." |
| Software Development | "I have an Xarray dataset that contains daily 3D temperature data. The dimensions are time, latitude, and longitude. Can you provide me with a Python function that subtracts the daily multi-year mean at each grid cell to obtain temperature anomalies?" |
| Software Development | "Here is my CV: ... Can you help me create a website in HTML format that shows my contact information and highlights my most relevant achievements? I want it to look simple and professional." |
| Software Development | "Please write a Python script for a cron job that will generate surface meteorological analyses over a 100 km × 100 km domain centered on Kansas City, Missouri every two hours." |

### *b. Generative AI limitations*

Limitations associated with generative AI apply across all student use cases and sometimes are related to task complexity. For instance, generative AI can struggle to create software for complex or uncommon tasks, such as computing a specific physical quantity that involves a series of steps or processing data from a non-standard format. In such cases, interactions with generative AI often require trial and error, as the AI occasionally repeats mistakes or overlooks errors previously corrected during the session. This can make the process tedious and time-consuming, sometimes leading to the conclusion that writing software without AI assistance might have been faster. These limitations can be ameliorated by providing sufficient context to generative AI and breaking down tasks into smaller, independent steps. While generative AI can provide valuable qualitative feedback, such as assessing tone or readability, fine-tuned tools specifically designed for text editing may offer more precise editing capabilities. Notably, content produced by generative AI may contain misleading or false information, which may negatively impact student learning and academic performance if not identified (Figure 2).

## 5. Ethical concerns of generative AI in education and research

### *a. Academic integrity*

The widespread adoption of generative AI has prompted schools, universities, and colleges to issue guidance addressing concerns related to academic integrity. Examples of Generative AI guidelines for teaching and research include those from the University of Maryland (https://ai.umd.edu/resources/guidelines) and Purdue University (https://www.purdue.edu/teaching-learning/instructors/ai.php). These guidelines establish institution-wide principles for collective responsibility surrounding the ethical and responsible use of generative AI in higher education. Not all institutions have yet implemented guidelines.

Reactions from educators have spanned the spectrum from full adoption of generative AI in their courses, provided its use is acknowledged, to a complete ban on its use. To facilitate transparency and a structured approach to integrating generative AI in the classroom, frameworks such as the AI Assessment Scale (Perkins et al. 2024) have been developed. Considerable underreporting of generative AI use by students persists[2] (Ling and Imas 2025), which may be ameliorated by such approaches. Students can also be invited to co-create academic integrity guidelines for courses, fostering a sense of ownership, responsibility, and rationale behind AI usage rules.

[2] Underreporting is due in part to desirability bias, which refers to underreporting of an action due to associated negative perceptions.

Assignments can be made more resistant to generative AI by using current, real-world examples whenever possible and mixing question types (e.g., equation derivation with its interpretation, or image interpretation with content synthesis). Students can also be required to document their thinking using citations or be directed to specific resources they should cite from, which allows them to practice referencing, literature review, and traceable science (i.e., explicitly citing information sources and documenting thought processes). Traceable composition tools, such as online synchronous documents, can also be used by students for assignments, as they will display writing and editing timestamps. Since definitions of academic misconduct can vary among educators and institutions, ensuring that academic misconduct is clearly defined and understood by students can help prevent accidental academic misconduct. Concrete examples of (in)appropriate use of generative AI can also help clarify the boundaries between academic misconduct and its allowed use in the classroom. Notably, AI content detection tools exhibit high false positives and should therefore be used with caution, if at all (Elkhatat et al. 2023). A classroom environment that incentivizes learning, teamwork, and practical skills, led by an educator who models desired behaviors concerning generative AI, may help navigate this paradigm shift in education.

Academic integrity is a responsibility shared by both students and educators. Accreditation frameworks exist to safeguard the quality of education, and this obligation extends to the adoption of emerging technologies. Although generative AI can assist in the creation of assessments and instructional activities, faculty oversight remains essential to ensure that such materials meet established academic standards. A scenario in which educators rely on AI to generate and grade work that students also complete with AI risks narrowing the scope of human judgment and diminishing the formative role of educators in the learning process, which historically has been an inherently human endeavor. At the same time, generative AI may also drive deeper engagement with knowledge, encouraging educators and students to sharpen critical thinking and use AI as a `collaborator' in discovery. Continued human-centered interpretation and inference of generative AI in education and research are crucial. Much like the internet and social media, generative AI may represent another pivotal shift in how knowledge is created and transmitted.

### *b. Technological inequities and transparency*

There are inequities and transparency issues regarding generative AI in education and research, which may increasingly require addressing head-on as advances in technology continue. Lack of transparency regarding training data and other details contributes to a lack of trust in generative AI. However, such information may be omitted due to safety implications and the competitive nature of the technology industry, as stated in OpenAI's ChatGPT-4 technical report (Achiam et al. 2023). The presence of misinformation, biases (Figure 3), and societal polarization in training datasets may propagate into content created by generative AI, as previously demonstrated (e.g., Luccioni et al. 2023), weakening science literacy and potentially compromising public safety. The leakage of biases can also occur in Earth systems applications, through biased physical science datasets used to train generative AI models or the models themselves (for examples, see McGovern et al. 2022). As generative AI tools become increasingly sophisticated, so will adversarial attacks, to which educators and students will be susceptible. The unguided or unstructured use of generative AI tools in the classroom could leak potentially sensitive or personal information into public databases, threatening student privacy. Established policies and guidelines may help mitigate such risks.

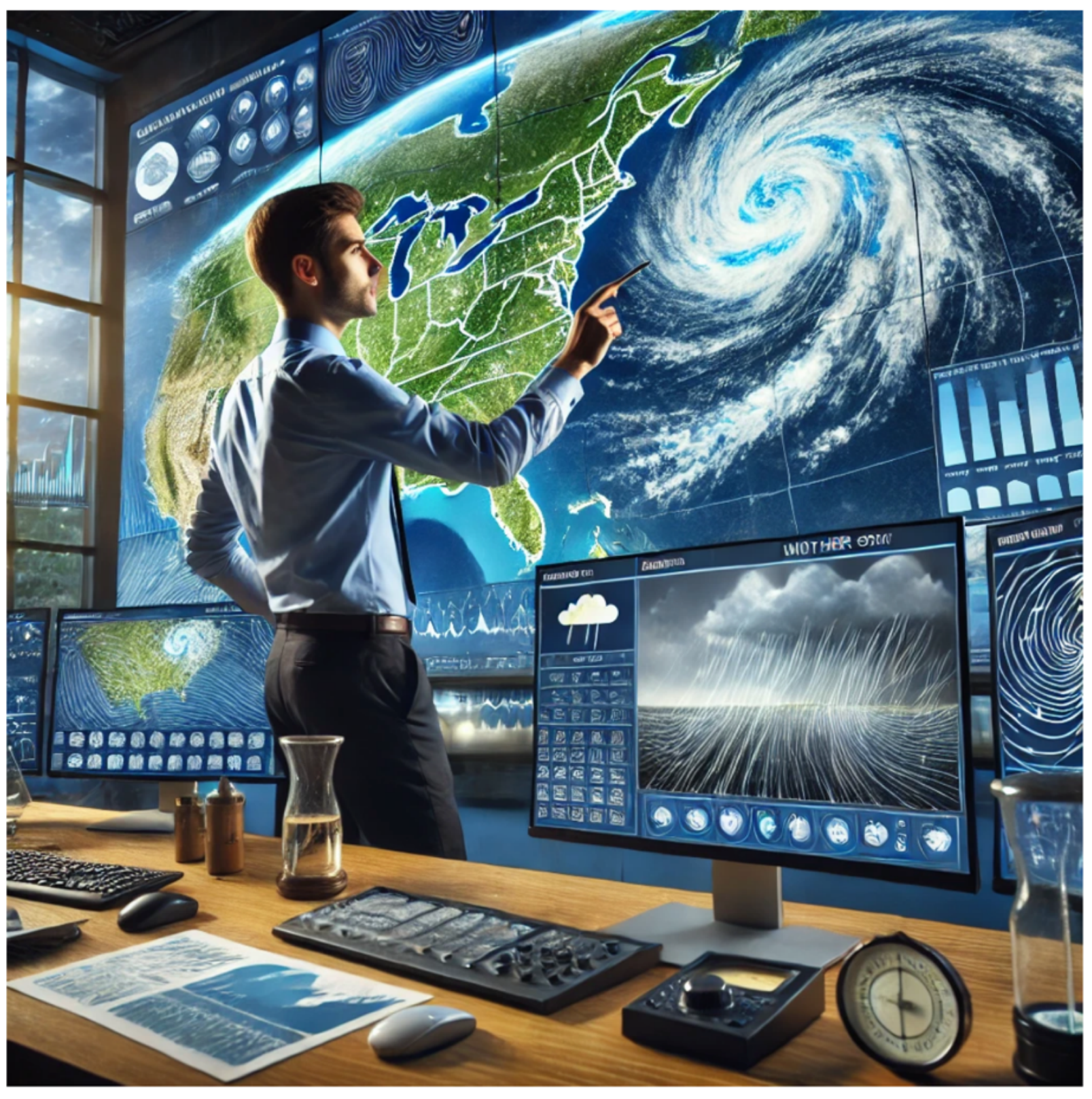

Figure 3. DALL-E 3 generated image from the prompt, ``Can you create an illustration of a meteorologist at work?" What underlying assumptions and biases do you see in the produced image?

There is also the possibility that generative AI could widen the inequities among different student populations. For example, paid versions and subscriptions that provide access to more advanced AI tools, or those that appear free initially but shift priorities to focus on profit, will offer reduced benefits for students and institutions with limited financial resources. Other inequities surround the language of choice for many large language models, which is Standard American English, disadvantaging those whose first language is not English. Students at non-R1 institutions may also lack suitable computing environments, further amplifying inequities. Yet it is these students, many of whom enter geoscience programs at regional, non-R1 institutions, often with gaps in their previous general education, who could benefit the most from individualized support provided by sophisticated AI tools. Faculty at non-R1 schools typically face heavy teaching loads, broad instructional responsibilities, and limited resources. Generative AI tools can benefit educators at these teaching-intensive institutions by reducing the time needed for course planning and material development. Efficiencies gained through the use of generative AI should be strategically redirected toward advancing learning and ensuring the continuous improvement of educational outcomes. Failure to do so risks undermining both instructional quality and the student experience at all institutions, but especially at non-R1 institutions where more in-person and collaborative student-teaching environments are expected. To harness the potential of AI equitably, we should incentivize open-source generative AI models (e.g., LLaMa; Touvron et al. 2023) or institutional investment in AI tools (see TerpAI by the University of Maryland: https://ai.umd.edu/resources/services), and promote adoption of generative AI especially at regional and non-R1 institutions.

*c. Energy and resource consumption*

As atmospheric scientists, educators should also address the energy use and carbon footprint of generative AI. By some estimates, multi-purpose generative AI tools use about ten times more energy than a search engine to execute a similar query during inference (Luccioni et al. 2023a). The power and water consumption of U.S. data centers is also considerable, with about a fifth relying on water from water-stressed watersheds (Siddik et al. 2021). While quantifying such estimates helps understand scale, the lack of technology transparency makes exact estimates across models difficult (Luccioni et al. 2024). Users should utilize generative AI when necessary or when it is highly beneficial for research and learning purposes. Less environmentally detrimental tools should be used when possible, such as those that employ sparsification techniques (Gale et al. 2019) or knowledge distillation (Jiao et al. 2019) , which are methods to reduce model complexity and improve efficiency without sacrificing much performance.

Not all is bad regarding the energy and resource consumption of generative AI. Once trained, GenCast can generate global numerical weather forecasts using a single Tensor Processing Unit in a few minutes, compared to the several hours and central processing units required for traditional physics-based systems (Price et al. 2023). Model architectures can also be developed that optimize energy efficiency, with recent efforts focusing on domain-specific hardware accelerators (Wang et al. 2025). AI-based tools also aid in resource and energy management, such as by more effectively integrating solar power into the energy grid, with some efforts noting benefits several orders of magnitude greater than their carbon cost (Dixon et al. 2022).

## 6. The path forward: Leveraging opportunity ethically

Institutions and professional organizations should strive to position educators at the forefront of this new paradigm, rather than merely responding to it, allowing for innovation in education. Updating assignments and policies at the pace of generative AI technological advances is time-consuming, resource-intensive, and complex. However, educators must stay informed about the current applications of generative AI in the workforce to prepare students for their next career steps. Tools like the "Generative AI Product Tracker" provide an example of an evolving comprehensive listing of generative AI tools for post-secondary education, and institutions are encouraged to curate and share such tools with faculty to support ongoing adaptation: https://sr.ithaka.org/our-work/generative-ai-product-tracker/. Institutions have a clear role to play in facilitating and incentivizing such professional development.

Addressing ethical concerns about generative AI requires widespread education, community organization, and collaboration, which could be potentially aided through open-source and communication platforms. Sharing resources and experiences with generative AI, as we do herein, can also help raise awareness and promote broader adoption of technological advances. As new versions of generative AI models are released, it is imperative that their fine-tuning be based on a diverse set of users, including users from non-R1 institutions. Based on the authors' teaching experiences, generative AI has helped accelerate workflows, create engaging classroom activities, and encourage a shift towards higher levels of Bloom's taxonomy that require more creativity, skepticism, and synthesis. Given the rapid adoption and widespread use of generative AI by students and the public, there is no going back; generative AI provides opportunities to engage with and produce information. It is up to us to find ways to ensure its ethical use and adoption.

*Acknowledgments.* This material is based upon work supported by the University of Maryland Grand Challenges Grants Program No. GC17-2957817 and the U.S. National Science Foundation under Grant No. ICER-2019758 and Grant No. 2425735. This piece was inspired by the participation of the authors on a panel about the same topic at the 2024 AMS Annual Meeting. We also thank Prof. Eduardo Araujo-Pradere from Miami Dade College, Dr. Karl Payne from the University of the West Indies, and Prof. Philippe Tissot from Texas A&M University-Corpus Christi for their perspectives on the use of generative AI in the classroom.

*Data availability statement.* No datasets were generated or analyzed in this manuscript.

**References**

Achiam, J., and Coauthors, 2023: GPT-4 technical report. Tech. rep., OpenAI. https://doi.org/10.48550/arXiv.2303.08774.

Abramson, J., and Coauthors, 2024: Accurate structure prediction of biomolecular interactions with AlphaFold 3. Nature, 630, 493–500, https://doi.org/10.1038/s41586-024-07487-w.

Bull, C., and A. Kharrufa, 2023: Generative artificial intelligence assistants in software development education: A vision for integrating generative artificial intelligence into educational practice, not instinctively defending against it. IEEE Software, 41 (2), 52–59, https://doi.org/10.1109/MS.2023.3300574.

Christenson, C. E., J. E. Martin, and Z. J. Handlos, 2017: A synoptic climatology of Northern Hemisphere, cold season polar and subtropical jet superposition events. J. Climate, 30 (18), 7231–7246, https://doi.org/10.1175/JCLI-D-16-0565.1.

Dixon, B., M. P´erez-Ortiz, and J. Bieker, 2022: Comparing the carbon costs and benefits of low-resource solar nowcasting. arXiv:2210.04554, https://doi.org/10.48550/arXiv.2210.04554.

Elkhatat, A. M., K. Elsaid, and S. Almeer, 2023: Evaluating the efficacy of AI content detection tools in differentiating between human and AI-generated text. International Journal for Educational Integrity, 19 (1), 17, https://doi.org/10.1007/s40979-023-00140-5.

Gale, T., E. Elsen, and S. Hooker, 2019: The state of sparsity in deep neural networks. arXiv preprint arXiv:1902.09574, https://doi.org/10.48550/arXiv.1902.09574.

Goodfellow, I., J. Pouget-Abadie, M. Mirza, B. Xu, D. Warde-Farley, S. Ozair, A. Courville, and Y. Bengio, 2014: Generative adversarial nets. Advances in Neural Information Processing Systems, 27, 139–144.

Gouia-Zarrad, R., and C. Gunn, 2024: Enhancing Students' Learning Experience in Mathematics Class through ChatGPT. International Electronic Journal of Mathematics Education, 19 (3), https://doi.org/10.29333/iejme/14614.

Jiao, X., Y. Yin, L. Shang, X. Jiang, X. Chen, L. Li, F. Wang, and Q. Liu, 2019: TinyBERT: Distilling BERT for natural language understanding. arXiv preprint arXiv:1909.10351, https://doi.org/10.48550/arXiv.1909.10351.

Ling, Y., and A. Imas, 2025: Underreporting of AI use: The role of social desirability bias. Available at SSRN, https://doi.org/10.2139/ssrn.5232910.

Liu, C., and S. Yang, 2024: Application of large language models in engineering education: A case study of system modeling and simulation courses. International Journal of Mechanical Engineering Education, 03064190241272728, https://doi.org/10.1177/03064190241272728.

Luccioni, A. S., S. Viguier, and A.-L. Ligozat, 2024a: Estimating the carbon footprint of BLOOM, a 176-billion parameter language model. Journal of Machine Learning Research, 24 (253), 11 990–12 004, https://doi.org/10.5555/3648699.3648952.

Luccioni, S., C. Akiki, M. Mitchell, and Y. Jernite, 2023: Stable bias: Evaluating societal representations in diffusion models. Advances in Neural Information Processing Systems, https://proceedings.neurips.cc/paperfiles/paper/2023/file/b01153e7112b347d8ed54f317840d8af-Paper-DatasetsandBenchmarks.pdf

Luccioni, S., Y. Jernite, and E. Strubell, 2024b: Power hungry processing: Watts driving the cost of AI deployment? Proceedings of the 2024 ACM conference on fairness, accountability, and transparency, 85–99, https://doi.org/10.1145/3630106.3658542.365

Matzakos, N., and M. Moundridou, 2025: Exploring Large Language Models Integration in Higher Education: A Case Study in a Mathematics Laboratory for Civil Engineering Students. Computer Applications in Engineering Education, 33 (3), e70 049, https://doi.org/10.1002/cae.70049.368.

McGovern, A., I. Ebert-Uphoff, D. J. Gagne II, and A. Bostrom, 2022: Why we need to focus on developing ethical, responsible, and trustworthy artificial intelligence approaches for environmental science. Environmental Data Science, 1, e6, https://doi.org/10.1017/eds.2022.5.

McMurtrie, B., 2024: Teaching: When AI is everywhere, what should instructors do next? The Chronicle of Higher Education.

Molina, M. J., and Coauthors, 2023: A review of recent and emerging machine learning applications for climate variability and weather phenomena. Artificial Intelligence for the Earth Systems, 2 (4), https://doi.org/10.1175/AIES-D-22-0086.1.239.

Nam, D., A. Macvean, V. Hellendoorn, B. Vasilescu, and B. Myers, 2024: Using an LLM to help with code understanding. Proceedings of the IEEE/ACM 46th International Conference on Software Engineering, 1–13, https://doi.org/10.1145/3597503.3639187.

Nancekivell, S. E., P. Shah, and S. A. Gelman, 2020: Maybe they're born with it, or maybe it's experience: Toward a deeper understanding of the learning style myth. Journal of Educational Psychology, 112 (2), 221–235, https://doi.org/10.1037/edu0000366.242.

Pardos, Z. A., and S. Bhandari, 2024: ChatGPT-generated help produces learning gains equivalent to human tutor-authored help on mathematics skills. Plos one, 19 (5), e0304 013, https://doi.org/10.1371/journal.pone.0304013.

Perkins, M., L. Furze, J. Roe, and J. MacVaugh, 2024: The Artificial Intelligence Assessment Scale (AIAS): A framework for ethical integration of generative AI in educational assessment. Journal of University Teaching and Learning Practice, 21 (6), 49–66, https://doi.org/10.53761/q3azde36.

Price, I., and Coauthors, 2023: Gencast: Diffusion-based ensemble forecasting for medium-range weather. arXiv preprint arXiv:2312.15796, https://doi.org/10.48550/arXiv.2312.15796.

Puryear, B., and G. Sprint, 2022: GitHub copilot in the classroom: Learning to code with AI assistance. Journal of Computing Sciences in Colleges, 38 (1), 37–47, https://doi.org/10.5555/3575618.3575622.245.

Ravuri, S., and Coauthors, 2021: Skillful precipitation nowcasting using deep generative models of radar. Nature, 597 (7878), 672–677, https://doi.org/10.1038/s41586-021-03854-z

Rombach, R., A. Blattmann, D. Lorenz, P. Esser, and B. Ommer, 2022: High-resolution image synthesis with latent diffusion models. Proceedings of the IEEE/CVF conference on computer vision and pattern recognition (CVPR), 10 684–10 695, https://doi.org/10.48550/arXiv.2112.10752.

Rumelhart, D. E., G. E. Hinton, and R. J. Williams, 1986: Learning representations by back-propagating errors. Nature, 323 (6088), 533–536, https://doi.org/10.1038/323533a0.

Siddik, M. A. B., A. Shehabi, and L. Marston, 2021: The environmental footprint of data centers in the United States. Environmental Research Letters, 16 (6), 064 017, https://doi.org/10.1088/1748-9326/abfba1.

Touvron, H., and Coauthors, 2023: LLaMa: Open and efficient foundation language models. arXiv preprint arXiv:2302.13971, https://doi.org/10.48550/arXiv.2302.13971.

Vaswani, A., 2017: Attention is All You Need. arXiv preprint arXiv:1706.03762, https://doi.org/10.48550/arXiv.1706.03762.251

Wang, I., and Coauthors, 2025: Carbon Aware Transformers Through Joint Model-Hardware Optimization. arXiv preprint arXiv:2505.01386, https://doi.org/10.48550/arXiv.2505.01386.

Weizenbaum, J., 1966: ELIZA—a computer program for the study of natural language communication between man and machine. Communications of the ACM, 9 (1), 36–45, https://doi.org/10.1145/365153.365168.254

Yau, M. K., and R. R. Rogers, 1996: A short course in cloud physics. Elsevier.

Zhai, C., S. Wibowo, and L. D. Li, 2024: The effects of over-reliance on AI dialogue systems on students' cognitive abilities: a systematic review. Smart Learning Environments, 11 (1), 28, https://doi.org/10.1186/s40561-024-00316-7.